\documentclass[useAMS,usenatbib]{mn2e}

\usepackage{graphicx}
\usepackage{bm}
\usepackage{natbib}

\def\apj{ApJ}                 % Astrophysical Journal
\def\aap{A\&A}                % Astronomy and Astrophysics
\def\mnras{MNRAS}             % Monthly Notices of the RAS
\title[Corotating light cylinders and Alfv\'en waves]{Corotating light cylinders and Alfv\'en waves}
\author[]{K.N. Gourgouliatos $^{1,}$$^{2}$\thanks {E-mail: kgourgou@purdue.edu; dlb@ast.cam.ac.uk} and D. Lynden-Bell$^{1}$\footnotemark[1]
\\$^{1}$Institute of Astronomy, University of Cambridge, Madingley Road CB3 0HA
\\$^{2}$Department of Physics, Purdue University, 525 Northwestern Avenue, West Lafayette, IN 47907-2036, USA}
\begin{document}

\date{Accepted -. Received -; in original form -}
\pagerange{\pageref{firstpage}--\pageref{lastpage}} \pubyear{-}
\maketitle

\label{firstpage}
\begin{abstract}
Exact relativistic force free fields with cylindrical symmetry are explored. Such fields are generated in the interstellar gas via their connection to pulsar magnetospheres both inside and outside their light cylinders. The possibility of much enhanced interstellar fields wound on cylinders of Solar system dimensions is discussed but these are most likely unstable.
\end{abstract}

\begin{keywords}
MHD -- methods: analytical -- stars: magnetic field -- stars: neutron -- stars: pulsars.
\end{keywords}

\section{Introduction}
\label{introduction}

Despite the pioneering work of \cite{GJ1969} on pulsar magnetospheres created by a rotating aligned dipole and of \cite{M1973} on relativistic winds both of which have been further explored in \cite{M1999},  there are few exact analytic solutions in relativistic magnetohydrodynamics (MHD). Prendergast's (2005) solution for a point explosion of force-free magnetic energy was the foundation of our recent work on that theme \citep{GL2008,G2009,GV2010} but recent advances have been mainly computational. The early axisymmetric static solution by \cite{SW1973} has been followed by accurate computations of \cite{K2006} and \cite{G2004} which give solutions for the steady aligned pulsar while \cite{KC2009} considered both time-dependent instabilities at the edge of the corotating zone and the non-aligned case. \cite{S2006} studied the time dependent relativistic problem of both the aligned and the oblique rotator; in the case of the aligned rotator he confirmed the results of \cite{CKF1999} who have studied the time-independent problem, whereas for the oblique rotator he presented the three-dimensional structure of the magnetosphere. Recent evidence that the pulses are generated near the light cylinder reinforce ideas put forward by \cite{DK1982,DK1985} who considered the orthogonally aligned cylindrical pulsar. Numerical solutions more relevant to active galactic nuclei have been developed by \cite{G2003} and \cite{MN2007}.  More recently \cite{TMN2009} have studied the problem of a spinning monopole, and they found that the combined effect of rotation and the magnetic field provide sufficient conversion of magnetic energy to kinetic with profound applications to GRBs but also to pulsar systems.

Force-free fields have been used as an approximation for pulsar magnetospheres. Leaving a magnetic field configuration to relax, we take a final state of a force-free magnetic field. Taking into account that the pulsar magnetosphere is dominated by the electromagnetic fields and considering that the effects of inertia are weaker makes the force-free assumption a valid starting point. The non-relativistic force-free approximation requires that the electric current flows along the magnetic field lines but neglects any effect of the electric field on the charges. In relativistic MHD the electric field is no more negligible as the electric field, in non-resistive MHD, is given by the cross product of the magnetic field with the velocity scaled to the speed of light. Thus the relativistic force-free approximation requires that the force of the magnetic field on the currents and that of the induced electric field on the charges add to zero. 

Here we explore the simplest exact relativistic force-free fields, those with cylindrical symmetry, and their coupling to rotating magnetospheres which corotate out to the light cylinder. While these solutions are very limited they allow us to answer questions such as: When one end of a cylinder of force-free field is rotated  at relativistic speed, is its flux per unit area greater than or less than the static field that surrounds it in pressure equilibrium, i.e. if we start rotating one end of a tube of flux, does the twist generated in it lead to its expansion or does its cross-section remain the same or even contract? What if anything happens as the rotation velocity of the tube approaches the speed of light? How rapidly does the rotation advance along the tube? For non-relativistic MHD this problem was treated in detail in \cite{S1994} but inevitably that has little bearing on light cylinders. We are studying idealized MHD in which the effects of resistivity are absent, therefore field lines that join the source of twist rotate while those that do not are left static. Thus a discontinuity is no surprise. In Section 2 we give the basic equations and solve them to find the general steady cylindrical force-free fields. Section 3 explores some simple special cases in greater detail, while Section 4 gives the time-dependent exact solution for rotationally generated Alfv\'en waves traveling down a uniform field. Section 5 considers how a rotating star or a pulsar might generate such waves in the interstellar magnetic field. It also presents the conundrum of how a pulsar's field that extends beyond the light cylinder can connect to an anti-aligned interstellar field and suggests a possible resolution. 
            
\section{Basic Equations}

We take the magnetic field and the velocity of the fluid to lie on cylinders and in cylindrical polars $(R, \phi, z)$ so $\nabla \cdot \bm{B}=0$ and
\begin{eqnarray}
\bm{B}=[0, B_{\phi}(R), B_{z}(R)]\,,
\label{magnetic}
\end{eqnarray}
\begin{eqnarray}
\bm{v}=c\bm{V}=c[0, V_{\phi}(R), V_{z}(R)]\,. 
\label{velocity}
\end{eqnarray}
The fluid is perfectly conducting so 
\begin{eqnarray}
\bm{E}=-\bm{V} \times \bm{B}=[B_{\phi}V_{z}-B_{z}V_{\phi}, 0, 0]\,,
\label{electric}
\end{eqnarray}
since $E_{R}$ is independent of $\phi$ and $z$ it follows that $\nabla \times \bm{E}=0$ so $\bm{B}$ is time independent. The charge density, $\rho$, is given by 
\begin{eqnarray}
\nabla \cdot \bm{E}=4 \pi \rho=\frac{1}{R}\frac{\partial }{\partial R}[R(B_{\phi}V_{z}-B_{z}V_{\phi})]\,.
\label{charge}
\end{eqnarray}
Provided $\bm{V}$ is independent of time the current density, $\bm{j}$, is given by
\begin{eqnarray}
4 \pi \bm{j}=\nabla \times \bm{B}=-\frac{\partial }{\partial R}B_{z}\bm{\hat{\phi}}+\frac{1}{R}\frac{\partial}{\partial R}(RB_{\phi})\bm{\hat{z}}\,.
\label{current}
\end{eqnarray}
The force-free condition is 
\begin{eqnarray}
\rho \bm{E} +\bm{j} \times \bm{B}=0\,.
\label{force}
\end{eqnarray}
Using the expressions for $\rho$ and $\bm{j}$ above, this becomes
\begin{eqnarray}
\frac{1}{2R^{2}}\frac{\partial}{\partial R}[R(\bm{V} \times \bm{B})]^{2}-\frac{1}{2}\frac{\partial}{\partial R}B_{z}^{2}-\frac{1}{2R^{2}}\frac{\partial}{\partial R}(RB_{\phi})^{2}=0\,.
\label{force2}
\end{eqnarray}
We notice that only the component of $\bm{V}$ perpendicular to $\bm{B}$ plays any role in equation (\ref{force2}). It is then convenient to write $\bm{V}$ in the form $\bm{V}=c^{-1}\bm{\Omega(R)} \times \bm{R}+V_{B}(R)\bm{\hat{B}}$, where $\bm{\hat{B}}=\bm{B}/|\bm{B}|$ is the unit vector along $\bm{B}$. We note that $V_{B}$ is not the component of $\bm{V}$  along $\bm{B}$
which contains also $(\bm{\Omega}\times  \bm{R})\cdot \bm{B}/|c\bm{B}|$. Thus any of our velocity fields may be considered as a (differential) rotation with flow along $\bm{B}$ superposed. Since the the flow along $\bm{B}$ does not contribute to the $(\bm{V} \times \bm{B})^{2}$ term in (\ref{force2}) that equation becomes
\begin{eqnarray}
\frac{\partial}{\partial R}(RB_{\phi})^{2}=\frac{\partial}{\partial R}(c^{-1}R^{2}\Omega B_{z})^{2}-R^{2}\frac{\partial}{\partial R}(B_{z}^{2})\,.
\end{eqnarray}
Integrating the last term by parts and then dividing by $R^{2}$ we find
\begin{eqnarray}
B_{\phi}^{2}=<B_{z}^{2}>-B_{z}^{2}(1-(c^{-1}\Omega R)^{2})\,,
\label{Bf}
\end{eqnarray}
where $<B_{z}^{2}>=R^{-2}\int^{R^{2}}_{0}B_{z}^{2}dR^{2}$ is the area average of $B_{z}^{2}$ up to $R$. Evidently if $\Omega(R)$ and $B_{z}(R)$ are specified such that the right hand side of (\ref{Bf}) is non-negative, then (\ref{Bf}) determines $B_{\phi}$ (up to a sign). With $\bm{B}$ and $\bm{V} \times \bm{B}$ then known, $\bm{E}$ follows from (\ref{electric}), $\rho$ from (\ref{charge}) and $\bm{j}$ from (\ref{current}). Thus the whole electromagnetic configuration is determined but to find the flow we must also specify the velocity $V_{B}(R)$ along $\bm{B}$. This is subject to the restriction that $V^{2} \leq 1$ which gives $(R\Omega/c+V_{B}B_{\phi}/B)^{2}+V_{B}^{2}B_{z}^{2}/B^{2} \leq 1$ that is $V_{B}^{2}+2V_{B}(R \Omega/c)(B_{\phi}/B) \leq 1-R^{2}\Omega^{2}/c^{2}$, which can be satisfied by a real $V_{B}$ for example $V_{B}=-2\frac{R \Omega}{c}\frac{B_{\phi}}{B}$ provided $1-(\frac{R\Omega}{c})^{2}(\frac{B_{z}}{B})^2\geq 0$ which is always satisfied since from (\ref{Bf}) the expression on the left is $<B_{z}^{2}>/B^{2}$. Notice there is no restriction that $R \Omega/c \leq 1$ because the fluid can move along the lines of the force to counteract such a 'superluminal' rotation leaving its total $V<1$. This is true however large $R\Omega/c$ becomes!

\section{Simple special cases}

Looking at equation (\ref{Bf}) there are several special cases that are worth noticing because they illustrate situations of particular simplicity. All the fields so far discussed are helical so it is useful here to define the pitch $\kappa(R)$ of a helical field line. The equation of such a field line is given by   $R d\phi/B_{\phi} = dz/B_{z}$ with $R$ constant. The pitch is defined as $\kappa(R)= d\phi/dz$ so $\kappa R = B_{\phi}/B_{z}$ .

(i) If $B_{z}$ is independent of $R$ then $<B_{z}^{2}>=B_{z}^{2}$ so from (\ref{Bf}) $B_{\phi}^{2}=(\Omega R^{2}/c^{2})B_{z}^{2}$. When $\Omega R=c$, $B_{\phi}= \pm B_{z}$ but as pointed out above there is no restriction that $\Omega R < c$ so we can have situations with $|B_{\phi}|>|B_{z}|$ at large $R$. There is however a tendency for such solutions to be unstable. In all cases the lines of force are given by $R=$const and $\frac{Rd\phi}{B_{\phi}}=\frac{dz}{B_{z}}$. The pitch of the resultant helix is $\kappa=d\phi/dz$ so $\kappa R=B_{\phi}/B_{z}$.

(ii)Writing equation (\ref{Bf}) in terms of the pitch $\kappa(R)$ we find 
\begin{eqnarray} 
B_{z}^{2}[1+(\kappa^{2}-\frac{\Omega^{2}}{c^{2}})R^{2}]=<B_{z}^{2}>\,,
\label{Bz}
\end{eqnarray}
A remarkable special case occurs when $\kappa^2-\Omega^2/c^2$ is independent of $R$.  This occurs when for example both the pitch and the angular velocity are both constant. Then we may multiply equation (\ref{Bz}) by $R^{2}$ and differentiate the result with respect to $R^{2}$; this gives us $B_{z}^{2}$ on the right which cancels against a term on the left. Dividing the resultant
equation by $R^{2}$ yields
\begin{eqnarray}
\frac{dB_{z}^{2}}{dR^{2}}[1+(\kappa^{2}-\frac{\Omega^{2}}{c^{2}})R^{2}]+2(\kappa^{2}-\Omega^{2}/c^{2})B_{z}^{2}=0\,.
\end{eqnarray}
This integrates to give, on taking the square root
\begin{eqnarray}
|B_{z}(R)|=\frac{B_{z}(0)}{1+(\kappa^{2}-\Omega^{2}/c^{2})R^{2}}\,.
\label{Bz2}
\end{eqnarray}
Notice that the field is uniform if $\kappa^{2}=\Omega^{2}/c^{2}$. If now a rotating tube of force is surrounded by a static uniform field $B_{0}$ at equilibrium with it at $R=a$
\begin{eqnarray}
B^{2}_{\phi}+B^{2}_{z}+4 \pi \sigma E_{R}=B_{0}^{2}\,,
\end{eqnarray}
where $\sigma$ is the surface charge on the rotating cylinder and $E_{R}$ is the electric field component inside the boundary $R=a$. By Gauss's theorem $4 \pi \sigma=-E_{R}(a)$ and from (\ref{electric}) $E_{R}^{2}=(c^{-1}\Omega aB_{z})^{2}$. Hence $|B_{z}(a)|=B_{0}/F^{1/2}$ where $F=1+(\kappa^{2}-\Omega^{2}/c^{2})a^{2}$. Thus when $\kappa=\Omega/c$ the rotating uniform field has the same $|B_{z}|$ as the non-rotating field outside. More generally the flux along the cylinder of radius $a$ is
\begin{eqnarray}
\begin{array}{ll}
\int^{a}_{0}2 \pi B_{z}RdR & =\pi B_{z}(0)(\kappa^{2}-\Omega^{2}/c^{2})^{-1}\ln F  \\
                                 & =\pi B_{z}(a) (\kappa^{2}-\Omega^{2}/c^{2})^{-1}F \ln F  \\ 
                                 & = \pi a^2 B_{0}[(F-1)^{-1}F^{1/2}\ln F]\,.
\end{array}
\end{eqnarray}
This function of $F$ maximises at $\pi a^{2} B_{0}$ when $F=1$ that is when $\Omega= c \kappa$.  In all other cases the  equilibrium twisted field configuration carries less flux over the circle $R<a$ than the straight bounding field $B_0$. Thus although some feel that the twisting of the field should cause it to contract under the influence of the hoop stresses so generated, nevertheless this is more than offset by the increase in field strength per unit longitudinal flux which is also caused by the twisting of the
field. Notice however we have only proved this when $\kappa^{2} - \Omega ^{2}/c^{2}$ is independent of $R$ within the twisted region. However for $(\kappa^{2}-\Omega^{2}/c^{2})a^{2}$ between $-0.5$ and $+1.0$ the flux decrease is less than $2\%$ so there is only a $<1\%$ increase in the radius of the twisted rotating cylinder in such circumstances. An untwisted $\bm{B}$ field spinning about the axis of a cylinder whose top speed is $1/\sqrt{2}$ of the velocity of light only increases its radius by about about $1\%$!

(iii) Now suppose that the rotation is uniform out to $R=b$ but thereafter takes the Keplerian profile $\Omega=\Omega_{0}(b/R)^{3/2}$. We substitute this velocity profile in equation~(\ref{Bf}) 
\begin{eqnarray}
B_{\phi}^{2}=<B_{z}^{2}>-B_{z}^{2}(1-\Omega^{2}R^{2}/c^{2}) 
\label{Bf2}
\end{eqnarray}
Now the Alfv\'en waves generated by such a rotation will travel at the speed light in a force-free field (see section 4) so the pitch generated is given by $\kappa=d\phi/dz=-\Omega(R)dt/(cdt)$ so $B_{\phi}=B_{z}\Omega(R)R/c$ so the first and last terms of equation (\ref{Bf2}) cancel so we are left with $<B_{z}^{2}>=B_{z}^{2}$ so the z-component of the field is uniform despite its differential rotation i.e. we are back to (i). 

\section{Time dependent solutions}

Since the solutions that leave the flux through the cylinder unchanged have a uniform $B_{z}$ we now look for time dependent solutions of the form $\bm{B}=[0, B_{\phi}(R,z,t), B_{0}]$ so $\nabla \cdot \bm{B}=0$
\begin{eqnarray}
\bm{v}=c\bm{V}=[0,\Omega(R,z,t)R, 0]+V_{B} \bm{\hat{B}}\,,
\end{eqnarray}
then,
\begin{eqnarray}
\bm{E}=-\bm{V} \times \bm{B}=-\frac{\Omega}{c}B_{0}\bm{R}\,,
\end{eqnarray}
\begin{eqnarray}
\nabla \times \bm{E}=-B_{0}Rc^{-1}\frac{\partial \Omega}{\partial z}\bm{\hat{\phi}}=-\frac{1}{c}\dot{B_{\phi}} \bm{\hat{\phi}}\,,
\label{CurlE}
\end{eqnarray}
\begin{eqnarray}
\nabla \cdot \bm{E}=4  \pi \rho=-\frac{B_{0}}{cR}\frac{\partial}{\partial R}(\Omega R^{2})\,,
\end{eqnarray}
\begin{eqnarray}
\frac{1}{c}\bm{\dot{E}}=-\frac{\dot{\Omega}}{c^{2}}B_{0}\bm{R}\,,
\end{eqnarray}
\begin{eqnarray}
\nabla \times \bm{B}=-\frac{\partial B_{\phi}}{\partial z}\bm{\hat{R}}+\frac{1}{R}\frac{\partial}{\partial R}(RB_{\phi})\bm{\hat{z}}\,,
\end{eqnarray}
\begin{eqnarray}
\nabla \times \bm{B}-\frac{1}{c}\bm{\dot{E}}=(\frac{\dot{\Omega}}{c^{2}}B_{0}R-\frac{\partial B_{\phi}}{\partial z})\bm{\hat{R}}+ \nonumber \\
+\frac{1}{R}\frac{\partial }{\partial R}(R B_{\phi})\bm{\hat{z}}=4 \pi \bm{j}\,.
\end{eqnarray}
From the force-free condition (\ref{force}) we have
\begin{eqnarray}
\bm{\hat{R}}\Big[\frac{1}{2R^{2}}\frac{\partial}{\partial R}(B_{\phi}^{2}R^{2})-\frac{B_{0}^{2}}{2R^{2}}\frac{\partial}{\partial R}\Big(\frac{\Omega^{2} R^{4}}{c^{2}}\Big)\Big]+\nonumber \\
\Big(\frac{\dot{\Omega}}{c^{2}}B_{0}R-\frac{\partial B_{\phi}}{\partial z}\Big)(B_{\phi}\bm{\hat{z}}-B_{0}\bm{\hat{\phi}})=0\,.
\label{force3}
\end{eqnarray}
The $\bm{\hat{z}}$ and $\bm{\hat{\phi}}$ components of this equation are satisfied provided 
\begin{eqnarray}
\frac{\dot{\Omega}}{c^{2}}B_{0}R=\frac{\partial B_{\phi}}{\partial z}\,,
\end{eqnarray}
but by (\ref{CurlE})
\begin{eqnarray}
B_{0}R\frac{\partial \Omega}{\partial z}=\dot{B_{\phi}}\,,
\end{eqnarray}
hence
\begin{eqnarray}
\frac{1}{c^{2}}\ddot{\Omega}=\frac{\partial^{2} \Omega}{\partial z^{2}}\,,
\end{eqnarray}
which wave equation is satisfied by $\Omega=\Omega[R, (t-z/c)]$. Then $B_{\phi}=-B_{0}R\Omega/c$ and putting this back into (\ref{force3}) the $\bm{\hat{R}}$ component is satisfied identically, so we have an exact solution for any wave form $\Omega(t-z/c)$. Of course we could superpose any $\Omega(t+z/c)$ but to solve the problem of a perfectly conducting disc at $z=0$ forced to rotate at a given time dependent rate $\Omega(t)$ we need the wave that travels out from the origin. Then the magnetic field at $z>0$, $R<a$ rotates at $\Omega(t-z/c)$ and that at $z<0$, $R<a$ rotates at $\Omega(t+z/c)$, see Fig.~1. Thus we have Alfv\'en waves generated by a rotating disc. At $R=a$ there is a surface density of charge $4 \pi \sigma=\frac{1}{c}\Omega a B_{0}$; $\Omega a <c$, which accounts for the electric field outside $R=a$. The changing surface density of charge involves a sheet current along the cylinder given by integrating (\ref{force3}) across the discontinuity in $B_{\phi}$ at $R=a$ thus $4 \pi J_{z}=B_{0} a \Omega/c$ and $\frac{\partial J_{z}}{\partial z}+\frac{\dot{\sigma}}{c}=0$ as required by charge conservation. Note that the net charge of the rotating cylinder is zero, as the volume charge density given by (\ref{charge}) and the surface charge density when integrated over the cylindrical volume and the discontinuity surface respectively add up to zero, so the electric field outside the cylinder is zero. \cite{K2002} in the context of Blandford-Znajek mechanism has investigated similar solutions.

\begin{figure}
  \centering
 \includegraphics[width=0.5\textwidth]{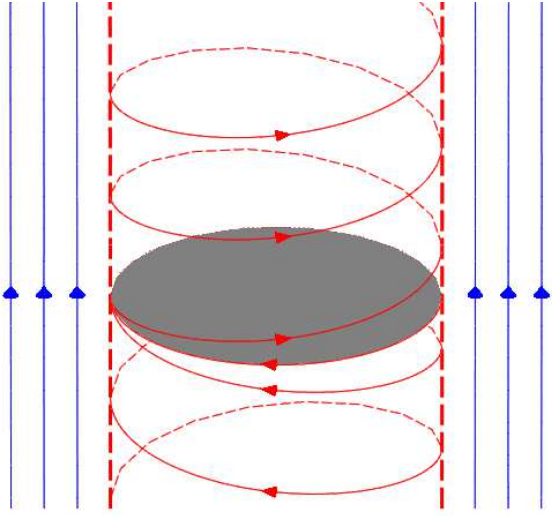}
     \caption{The magnetic field lines attached to a disc which is rotating rigidly. The twisted field is confined by a uniform external field. The heavy dashed line is the separatrix which carries a surface charge. When the tip speed of the disk is $c$ this is also the light cylinder.}
\end{figure}

\section{Rotating magnetic fields in the interstellar gas} 

So far our solutions have been exact but abstract. We now take a step nearer to the real world but the application is somewhat forced and the solution inexact.

Consider a star with a dipolar magnetic field with axis along $Oz$. Let it be embedded in an interstellar magnetic field oriented in the same direction. Then the field is as shown in Fig.~2 where the separatrix has been emphasised. The closed field lines near the star will lie in meridional planes and rotate with the star. The outermost field lines outside any separatrix will not rotate at all just like the outer field lines in Fig.~1. The remaining field lines that join the star to the interstellar gas will carry the torque that generates the Alfv\'en wave that propagates along the interstellar field to great distances from the star. In our force-free approximation these Alfv\'en waves travel at the velocity of light but in reality the fields will probably be too weak well away from the star so there the rotational disturbance will travel at the Alfv\'en speed given by $c_{A}^{2}=\frac{c^{2}B_{0}^{2}}{B_{0}^{2}+4 \pi \rho_{0}c^{2}}$ where $\rho_{0}$ is the gas density. In Fig.~2 we have assumed that the light cylinder lies beyond the asymptotic radius of the separatrix. For a strong stellar dipole or a weaker interstellar field this will not be true. We shall consider that situation afterwards. We write the field in the form
\begin{eqnarray}
\bm{B}=\frac{1}{2 \pi}(\nabla P \times \nabla \phi +\beta\nabla \phi)\,.
\end{eqnarray}
Then for a dipole $\bm{M}$ in a straight external field we find
\begin{eqnarray}
P=\pi(B_{0}r^{2}+2Mr^{-1})(1-\mu^{2})
\end{eqnarray}
The separatrix $P=P_{S}=3 \pi B_{0}^{1/3}M^{2/3}$ which for large $z$ gives $R_{S}=a=\sqrt{3}(M/B_{0})^{1/3}$. Thus all the flux with $P<P_{S}$ links the star to the interstellar magnetic field. At large $z$ we have from Section 3 (i) and under equation~(27) $\beta=-2\pi R^{2}B_{0}\Omega/c =-2\Omega P/c$, so since near the star we need $\beta$ to carry the torque away from the star we set $\beta=-2\Omega(t) P/c$ for $z \approx 0$, $P < P_{S}$ and $ \beta=-2\Omega(t-z/c)P/c$ for $z$ large since that gives the field found in Section 4. We therefore choose as our approximation $\beta=-2\Omega(t-z/c)P/c$ whenever $z>0$ and $P < P_{S}$. For $P > P_{S}$,  $\beta = 0$, see Fig~3. Since we have used the $P$ of the non-rotating configuration which will in practice be modified by the rotating charges this is only approximate; however we note that in our rotating cylinder the $B_{z}$ remained at $B_{0}$ even in the presence of the rotation so at large $z$ the proposed solution is exact. 
\begin{figure}
  \centering
 \includegraphics[width=0.43\textwidth]{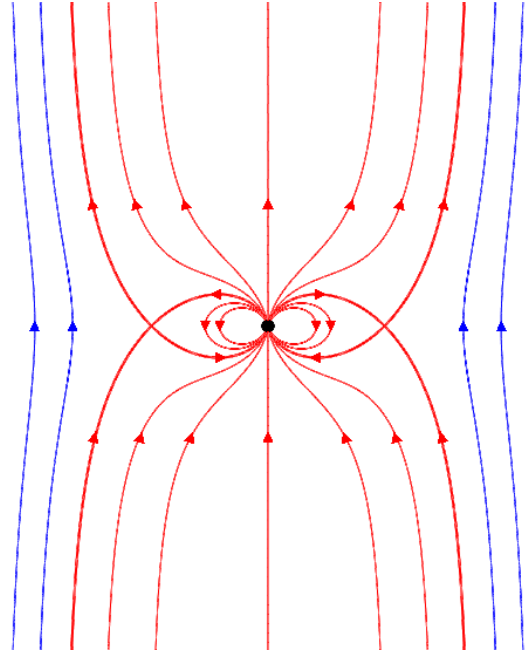}
     \caption{The magnetic field of a dipole embedded inside a uniform external magnetic field of the same polarity. The thick red line is the separatrix of the field lines that are attached to the dipole field and those detached. Note that amongst the attached field lines the ones emerging near the poles reach infinity whereas the ones emerging from the area near the equator close in the other hemisphere of the dipole.}
\end{figure}

Now we consider the case in which the light cylinder is of smaller radius than the asymptotic radius, $a$, of the separatrix of the static field. Close to the light cylinder the interstellar magnetic field is far too weak to have any effect so for the aligned pulsar, the work of \cite{GJ1969} as modified by \cite{SW1973} and most recently treated \cite{G2004} is appropriate. However in those treatments the far field has a weak poloidal component and spirals off to infinity with that component not far from radial. There the field will be radically modified if the pulsar resides in an asymptotic field from the interstellar medium because asymptotically the field from the pulsar must align along the cylinders and produce Alfv\'en waves propagating along them. $\Omega R$ will be much greater than $c$. Fig.~4 illustrates the type of the field configuration that we envisage. All the pulsar field lines that do not close within the light cylinder end up winding around the interstellar field and their $\Omega$ is still that of the pulsar!
\begin{figure}
  \centering
 \includegraphics[width=0.5\textwidth]{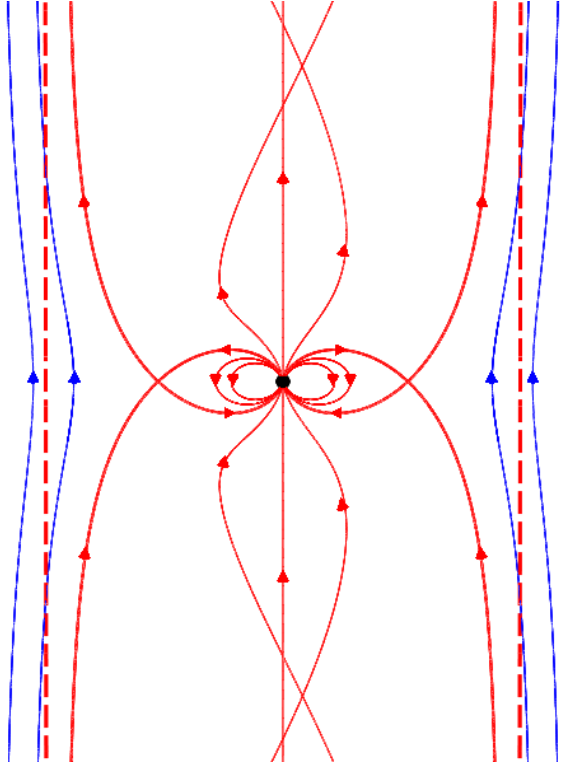}
     \caption{The magnetic field of a rotating dipole embedded inside a uniform external magnetic field of the same polarity. The discontinuity of the toroidal component of the magnetic field on the separatrix leads to the formation of a current sheet. We have assumed that the combination of the pulsar magnetic field, insterstellar magnetic field and rotation velocity is such that the light cylinder occurs outside the co-rotating loops. It is possible that the light cylinder while being outside the co rotating loop intersects with the separatrix,  in which case some of the pulsar's twisted field lines will pass through the light cylinder.}
\end{figure}
\begin{figure}
  \centering
 \includegraphics[width=0.5\textwidth]{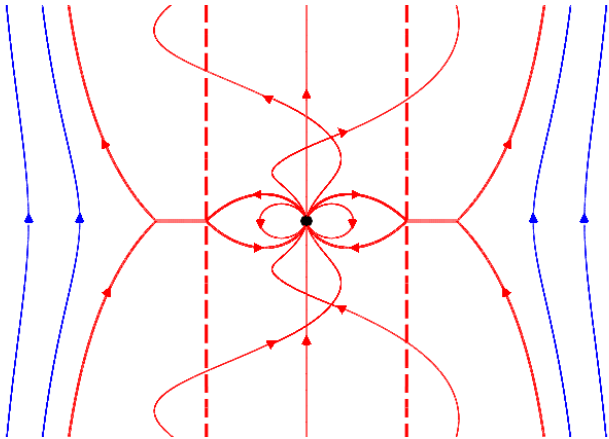}
     \caption{The magnetic field of a rotating dipole embedded inside a uniform external magnetic field of the same polarity. The discontinuity of the toroidal component of the magnetic field on the separatrix leads to the formation of a current sheet. Unlike Fig.~3 the combination of the fields and rotation velocity is such that the light cylinder occurs near the star. In this case a sheet current forms on the equator.}
\end{figure}
\begin{figure}
  \centering
 \includegraphics[width=0.5\textwidth]{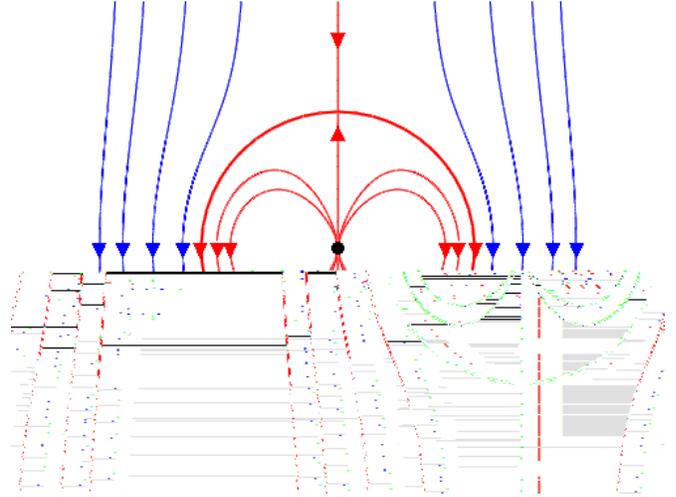}
     \caption{The magnetic field of a dipole embedded inside a uniform external magnetic field of the opposite polarity. The attached field lines to the dipole are enclosed within a spherical surface. The external field lines are deformed from being uniform close to the equator.}
\end{figure}
\begin{figure}
  \centering
 \includegraphics[width=0.5\textwidth]{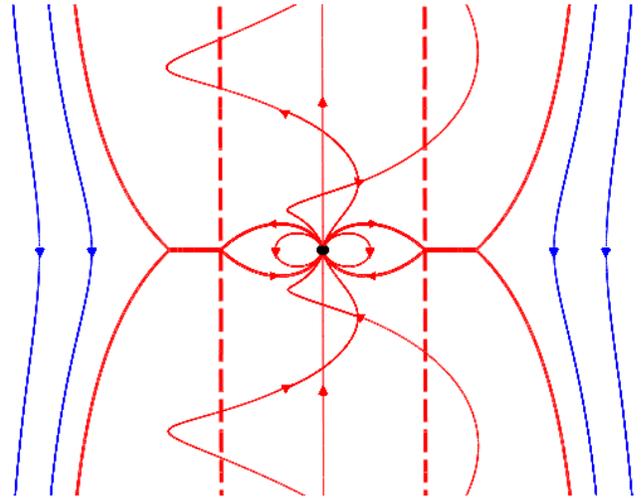}
     \caption{The magnetic field of a rotating dipole embedded inside a uniform magnetic field of the opposite polarity. The field's configuration is similar to that of Fig.~4, however the direction of the field lines is different. There is a discontinuity on the separatrix of the poloidal component of the magnetic field enhancing the current sheet that already exists because of the discontinuity on the toroidal component of the field.}
\end{figure}

The flux function for the pure dipolar field is $P = 2 \pi M (1 - \mu^2)/r$ so the flux that fails to reach the equator within the light cylinder in such a field would be $P_{c} ~= 2 \pi M \Omega/c$. In reality the field is distorted by the rotation to be more like $P = 2 \pi M/r(1 + (1/2)(r/r_c)^3)(1 - \mu^2)$  for $r < r_c = c/\Omega$
which gives $P_c = 3 \pi M \Omega/c$. A very rough model that gets the fluxes right (but is not force-free) is given by the $P$ above for $r < r_c$, $P = P_c (1-\mu^2)$, for $r_c < r < r_s$ where $r_{s}^2 = P_c/(3 \pi B_0)$ and $P = \pi B_0 r^2 ( 1 + 2 (r_s/r)^3)$  for $r > r_s$. As before the toroidal field is given as in equation~(28) with $\beta(P)= -2 \Omega(t-|z|/c)P/c$ for $P < P_c$ and zero elsewhere.

To see the orders of magnitude consider a pulsar whose polar field is $B_{p}$ and suppose the radius of the polar cap whose field does not close within the light cylinder is $R_{p}$ and the inter-stellar field is $B_{0}$. Then the asymptotic radius of the field attached to the pulsar will be $a=R_{p} \sqrt{B_{p}/B_{0}}$ suitable values might be $B_{p}=3 \times 10^{12} \rm{G}$, $B_{0} =3\times10^{-6} \rm{G}$ and $R_{p} =1\rm{km}$ giving $a= 10^{14}\rm{cm}$. The value of $R_{p}$  is linked to the pulsars rotation rate $\Omega$ and radius $r_{s}$ by $R_{p}= r_{s} \sqrt{r_{s} \Omega/c}$ which gives $R_{p} =1\rm{km}$ for $r_{s}=11\rm{km}$ for a pulsar with a period of $28\rm{ms}$. More generally $a=r_{s} \sqrt{(\Omega r_{s}/c)B_{p}/B_{0}} \propto T^{-1/2}$ where $T$ is the pulsar period. For the numbers above but general $T$, we find $\Omega a =3.5 \times 10^{3} c (T/\rm{s})^{-3/2}$ which is many times the speed of light. If the Alfv\'en wave advances at c then in the interstellar gas $B_{\phi}/B_{0}=3.5 \times 10^{3} (T/\rm{s})^{-3/2}$ in the idealised model and this is increased by the factor $c/c_{A}$ if the Alfv\'en speed is $c_{A}$. However such highly wound magnetic fields are likely to be highly unstable. In the ideal case the plasma has to move along the field so that $v_{z}= (B_{0}/B_{\phi}) c_{A}$. The field winds through the plasma like an Archimidean screw winds up the water. While these estimates will be complicated by instabilities they do indicate that pulsars generate much enhanced fields in the interstellar medium albeit over tubes of force with diameters similar to that of the solar system.

        A different but interesting situation occurs when the interstellar field is aligned oppositely to magnetic moment of the pulsar. The total flux of the pulsar outside its light cylinder, $P_c$, will be unable to counteract the flux of the interstellar field equatorially all the way to infinity since the latter grows like $R^{2}$. Furthermore the pulsar's flux can not cross its equatorial plane outside the light cylinder as by symmetry it would have to cross perpendicularly and the plasma attached would then move faster than light!However the flux from the pulsar has to reach infinity since it carries the flow of plasma pushed by the
Archimedian screw of the whirling field. The only place where the flux from the pulsar can deform the field over large distances is close to the
axis, indeed in the non-rotating case there would be a neutral point on axis \citep{D1958}, see Fig.~5. As the pulsar rotates the equatorial diameter of the field structure around it grows and this neutral point is sent to infinity along the axis. An oppositely directed poloidal field is generated near the axis and is accompanied by the $B_{\phi}$ and the plasma flow due to the pulsar's rotation. We illustrate both the weak spin case when all the star's field lies within the light cylinder and rotates with the star Fig.~5 and the pulsar case just described in Fig.~6.  Corresponding to any solution such as that shown in Fig.~4. there is an other solution with the field reversed outside the separatrix. If the longitudinal field inside the separatrix is small compared to the toroidal field there, the field pressures and tension still balance along the separatrix. The argument given above leads us to believe that this is the field configuration of the steady anti-aligned pulsar.

\section{Conclusions}

In this paper we have studied analytical solutions of rotating magnetic fields emerging from dipoles or discs embedded in uniform magnetic fields. This study is based on the assumption that the dipole field is either aligned or counter-aligned to the external field. Fields at an angle $i$ to the dipole may be considered by taking the solutions above with $B_0$ replaced by $B_0 \cos i$ and superposing a lateral field parallel to the plane of the disk of magnitude $B_0 \sin i$,  but this gives gives only orders of magnitude and is no substitute for a full non-axially symmetric treatment.

We find that the spinning field will form Alfv\'en waves in the interstellar gas. There is no need to confine these Alfv\'en waves inside the light cylinder, as the plasma can have a velocity component parallel to the magnetic field and not exceed the speed of light and we find that for typical pulsar values they will occupy cylinders whose cross sectional areas are of the size of the solar system. These field configurations with tightly wound fields outside the light cylinder are probably very unstable, in particular the pressure of the whole tube along the axis is very large so any departure from axial symmetry will be enhanced by an instability analogous to the fire-hose instability. However there remains the interesting possibility that the magnetic fields generated over small regions by the thrashing field lines might be surprisingly large although smaller than the maxima estimated via these axially symmetric models.

{}

\end{document}